\documentclass[12pt,a4paper]{article}
\usepackage[fleqn]{amsmath}
\usepackage{amssymb}

\newcommand{\dsum}{\sum}
\newcommand{\tsum}{\sum}
\newcommand{\dint}{\int}
\newcommand{\tciLaplace}{\mathcal{L}}

\newcommand{\be}{\begin{equation}}
\newcommand{\ee}{\end{equation}}
\newcommand{\bea}{\begin{eqnarray}}
\newcommand{\eea}{\end{eqnarray}}

\makeatletter
\def\section{\@startsection{section}{1}
{\z@}{-3.5ex plus -1ex minus -.2ex}{2.3ex plus .2ex}{\normalsize \bf}}
\def\subsection{\@startsection{subsection}{2}
{\z@}{-3.25ex plus -1ex minus -.2ex}{1.5ex plus .2ex}{\normalsize \sl}}
\def\subsubsection{\@startsection{subsubsection}{3}
{\z@}{-3.25ex plus -1ex minus -.2ex}{1.5ex plus .2ex}{\normalsize}}

\def\thedoctitle{\bf
MaxEnt Mechanics
}

\def\theauthorname{
Jean-Bernard Brissaud
}

\def\authoraddress{
Lab/UFR High Energy Physics, Physics Department, Faculty of Sciences, Rabat, Morocco.
\\ E-mail: jb.brissaud@gmail.com
}

\def\thereceivedhistory{10 January 2007}

\begin{document}
\parindent 0cm

{ \noindent {\\ \\} }

{ \LARGE \noindent \thedoctitle } \\

{ \noindent {\bf {\theauthorname}} } \\

{ \normalsize \noindent {\authoraddress} } \\

{ \noindent {\sl \thereceivedhistory} } \\

\vspace{2pt} \hbox to \hsize{\hrulefill}
\vspace{.1in}

\noindent

{\bf Abstract:}
This paper shows that: (a) given a mechanical system described by a set of
independent coordinates in configuration space, (b) given an initial state
of specified initial coordinates, and (c) given a situation in which the
system can follow any one of a set of different possible paths with a
pertinent probability $p_{i}$ , then the average path (defined as the
weighted average positions and corresponding times) will obey Lagranges'
equations iff the Shannon information defined by the distribution of
probabilities is an extreme (maximum) compared to any imaginable other
distribution. Moreover, the resulting action function is proportional to
this Shannon information.
\\

\\

\vspace{2pt} \hbox to \hsize{\hrulefill}

\newpage

\def\oddmarkC{\thepage}
\def\oddmarkD{}
\def\oddmarkE{}
\def\oddmarkF{}

\section*{Introduction}

Physics is fundamentally divided in two parts. Mechanics, which includes
classical, relativistic and quantum mechanics, is ruled by the least-action
principle (LAP) and characterized by laws invariant by time reversal.
Thermodynamics, on the other hand, can be entirely deduced from the maximum
entropy principle (Maxent), and introduces a so-called "arrow of time".

LAP is equivalent to Newton's law in classical mechanics, and can be used in
relativistic and quantum mechanics with an adequate reformulation of the
Lagrangian. It has been historically applied, from Fermat to Feynmann, with
such success that today it is a fundamental principle of physics. It has
never been, to our knowledge, deduced from more fundamental principles. On
the other hand, Maxent is justified by the Bayesian rules of probability,
and simply gives the most probable probability distribution according to
given constraints. For instance, in equilibrium thermodynamics, Maxent
proves that the average values of the extensive quantities used to describe
the system (energy, volume, ...) are appropriate constraints. For an
excellent summary of Maxent properties and their applications to
thermodynamics, see \cite{Dewar}. Since Maxent is a principle of logic, it
can and has been successfully applied in many different fields, for example
biology, economy or ecology\cite{growLn}\cite{Biophysics}.

A mechanical system is defined by the values of a set of generalized
coordinates at a given time. We make the hypothesis that these values,
including time, are only known in average. Applying Maxent, we will
logically deduce that the system will follow the Lagrangian laws of motion.
LAP is a consequence of Maxent with appropriate constraints.

\bigskip

In the first section, a least information principle will be deduced from
Maxent. This section does not involve physics and is entirely mathematical.

In the second section, the least information principle will be applied to a
mechanical system whose space and time coordinates are only known in
average. The least information principle will appear to be, up to a
multiplicative constant with dimension $Action$, the least action principle.

A third section will be devoted to possible \ physical interpretations of
this result, and to new questions that then naturally emerge.

\section{The least information principle}

In this section, a least information principle, similar in form to the least
action principle, will be deduced from Maxent applied with linear
constraints. Provided that time is one of the constrained quantities, the
rate of information will satisfy the equations of Cauchy-Riemann.\bigskip

\subsection{The proof}

Notations: we write a sequence $(s_{i})$ instead of $(s_{1},s_{2},...,s_{i_{%
\max }})$. Two sequences using the same name of indice should have the same
length. We write $\tsum\limits_{i}$ instead of $\tsum\limits_{i=1}^{i_{\max
}}$. A sequence using two indices is written $(s_{i,j})$ instead of $%
(s_{1,1},...,s_{1,j_{\max }},...,s_{i,1},...,s_{i,j_{\max }},...,s_{i_{\max
},1},...,s_{i_{\max },j_{\max }})$.\bigskip

Some numbers $(A_{k,i})$ are given once and for all, and will always be
considered fixed. Some other numbers $(A_{k})$ are given and we wish to find
the probability law $(p_{i})$ satisfying the constraints:

\begin{equation}
\left\{ 
\begin{array}{l}
\dsum\limits_{i}p_{i}A_{k,i}=A_{k}\qquad for~all~k \\ 
\dsum\limits_{i}p_{i}=1%
\end{array}%
\right.  \label{system}
\end{equation}

Following Jaynes, we will choose the probability law which maximizes the
information:

\begin{equation*}
I=-\dsum\limits_{i}p_{i}\ln p_{i}
\end{equation*}

Using Lagrange multipliers $(\alpha _{k})$ (see demonstration in Annex 1),
we obtain the standard results:

\begin{equation}
dI=\dsum\limits_{k}\alpha _{k}dA_{k}  \label{dI}
\end{equation}

\begin{equation}
\frac{\partial \alpha _{k}}{\partial A_{l}}=\frac{\partial \alpha _{l}}{%
\partial A_{k}}\ \ \ \ for~all~k,l  \label{exactdiff}
\end{equation}

Adding another constraint on a new variable $t$ that the reader can
conveniently consider as time, the system (\ref{system}) then becomes:

\begin{equation}
\left\{ 
\begin{array}{l}
\dsum\limits_{i}p_{i}A_{k,i}=A_{k}\qquad for~all~k \\ 
\dsum\limits_{i}p_{i}t_{i}=t \\ 
\dsum\limits_{i}p_{i}=1%
\end{array}%
\right.  \label{system1}
\end{equation}

Calling $\beta $ the Lagrange multiplier of $t$, the results (\ref{dI}) and (%
\ref{exactdiff}) become:

\begin{equation}
dI=\dsum\limits_{k}\alpha _{k}dA_{k}+\beta dt  \label{dI1}
\end{equation}

\begin{equation}
\begin{tabular}{l}
$\frac{\partial \alpha _{k}}{\partial A_{l}}=\frac{\partial \alpha _{l}}{%
\partial A_{k}}\ \ \ \ for~all~k,l$ \\ 
$\frac{\partial \alpha _{k}}{\partial t}=\frac{\partial \beta }{\partial
A_{k}}$%
\end{tabular}
\label{exactdiff1}
\end{equation}

The notation $\overset{.}{A}_{k}$ is an abbreviation for the exact
differential $\frac{dA_{k}}{dt}$. Equation (\ref{dI1}) can be written

\begin{equation*}
dI=\left( \dsum\limits_{k}\alpha _{k}\overset{.}{A}_{k}+\beta \right) dt
\end{equation*}

Let%
\begin{equation}
L((A_{k}),(\overset{.}{A}_{k}),t)=\dsum\limits_{k}\alpha _{k}\overset{.}{A}%
_{k}+\beta  \label{L}
\end{equation}

We have:%
\begin{equation}
dI=Ldt  \label{dI_L}
\end{equation}

We now prove our main result, which is that $L$ satisfies the equations of
Cauchy-Riemann.

Since $\frac{\partial \overset{.}{A_{l}}}{\partial A_{k}}=0$ (but note that $%
\frac{\overset{.}{dA_{l}}}{dA_{k}}$ may be $\neq 0$), we have from equation (%
\ref{L})

\begin{equation}
\frac{\partial L}{\partial A_{k}}=\dsum\limits_{l}\frac{\partial \alpha _{l}%
}{\partial A_{k}}\overset{.}{A_{l}}+\frac{\partial \beta }{\partial A_{k}}
\label{L1}
\end{equation}

The $(\alpha _{k})$ are only functions of the $(A_{k})$ and $t$, $\frac{%
\partial \alpha _{l}}{\partial \overset{.}{A_{k}}}=\frac{\partial \beta }{%
\partial \overset{.}{A_{k}}}=0$ for all $k,l$. So we have:

\begin{equation}
\frac{\partial L}{\partial \overset{.}{A_{k}}}=\alpha _{k}  \label{dL0}
\end{equation}

and

\begin{equation}
\frac{d}{dt}\frac{\partial L}{\partial \overset{.}{A_{k}}}=\frac{d\alpha _{k}%
}{dt}=\dsum\limits_{l}\frac{\partial \alpha _{k}}{\partial A_{l}}\overset{.}{%
A_{l}}+\frac{\partial \alpha _{k}}{\partial t}  \label{dalpha0}
\end{equation}

As a result of (\ref{exactdiff1}),

\begin{equation*}
\frac{d}{dt}\frac{\partial L}{\partial \overset{.}{A_{k}}}=\dsum\limits_{l}%
\frac{\partial \alpha _{l}}{\partial A_{k}}\overset{.}{A_{l}}+\frac{\partial
\beta }{\partial A_{k}}
\end{equation*}

and, using (\ref{L1}), $L$ satisfies the equations of Cauchy-Riemann:

\begin{equation}
\frac{d}{dt}\frac{\partial L}{\partial \overset{.}{A_{k}}}=\frac{\partial L}{%
\partial A_{k}}\qquad for~all~k  \label{cauchy_riemann}
\end{equation}

\bigskip

As a consequence, the information

\begin{equation}
I=\dint\limits_{t_{0}}^{t_{1}}L(t)dt  \label{lip}
\end{equation}

is stationary. For any variations $\delta A_{k}(t)$ such that $\delta
A_{k}(t_{0})=\delta A_{k}(t_{1})=0$ for all $k$, we have $\delta I=0$.

\bigskip

By analogy with LAP, this consequence will be subsequently referred as the
"least information principle".

\subsection{Conservation laws}

From equations (\ref{dalpha0}) and (\ref{cauchy_riemann}), we obtain

\begin{equation}
\frac{d\alpha _{k}}{dt}=\frac{\partial L}{\partial A_{k}}\hspace{3cm}%
for~all~k  \label{dalpha}
\end{equation}

One can also prove that (see demonstration in Annex 2)

\begin{equation}
\frac{d\beta }{dt}=\frac{\partial L}{\partial t}  \label{dbeta}
\end{equation}

\bigskip

If $L$ does not depend explicitly on a given $A_{k}$, then the conjugate
quantity $\alpha _{k}$ does not vary with $t$. If $L$ does not depend
explicitly on time $t$, then the quantity $\beta $ does not vary with $t$.

\subsection{Degree of validity}

Note that in this problem, there are no assumptions about the nature of the $%
(A_{k})$. In particular:

- The $(A_{k})$ do not have to be frequency averages of the values $%
(A_{k,i}) $ in an experiment. Neither do we need a notion of ensemble. We
want to find a probability law which reflects our state of knowledge, not a
property of some system.

- The $(A_{k})$ do not have to be extensive quantities (a thermodynamical
concept not necessary for Maxent), neither do they need to scale together or
have any other relationship.

\section{Physical application}

Let us consider a mechanical system whose state is defined by a set of
independent coordinates $(q_{k})$. The motion of such a system can be
described by a parameterized curve $(q_{k}(\lambda ),t(\lambda ))$ in the $%
((q_{k}),t)$ space. We call such a curve a path. The system can potentially
take many different paths from a given starting position $(q_{k}(0),t(0))$.
Let us denote by $(i)$ the set of all these paths, and adopt, without loss
of generality, a common parameter $\lambda $ to describe all these paths. A
given path $i$ is then described by the $k+1$ functions $((q_{k,i}(\lambda
)),t_{i}(\lambda ))$.

We make the hypothesis that the observed path $((q_{k}(\lambda )),t(\lambda
))$ is the average path of all the paths $i$, each one occurring with a
probability $p_{i}$. Mathematically:

$\hspace{3cm}$%
\begin{equation*}
\left\{ 
\begin{array}{l}
\dsum\limits_{i}p_{i}q_{k,i}(\lambda )=q_{k}(\lambda )\qquad for~all~k \\ 
\dsum\limits_{i}p_{i}t_{i}(\lambda )=t(\lambda ) \\ 
\dsum\limits_{i}p_{i}=1%
\end{array}%
\right. \hspace{3cm}for~all~\lambda
\end{equation*}

We now fix the parameter $\lambda $, and no longer write the dependence on $%
\lambda $. The Maxent distribution which satisfies the preceding constraints
is the solution of system (\ref{system1}), with:

\begin{equation*}
\begin{tabular}{l}
$q_{k,i}=A_{k,i}$ \\ 
$q_{k}=A_{k}$%
\end{tabular}%
\end{equation*}

Equations (\ref{L}), (\ref{dI_L}), (\ref{dalpha}) and (\ref{dbeta}) become:

\begin{equation}
\begin{tabular}{l}
$L((q_{k}),(\overset{.}{q}_{k}),t)=\dsum\limits_{k}\alpha _{k}\overset{.}{q}%
_{k}+\beta $ \\ 
$dI=\dsum\limits_{k}\alpha _{k}dq_{k}+\beta dt=Ldt$ \\ 
$\frac{d\alpha _{k}}{dt}=\frac{\partial L}{\partial q_{k}}\hspace{3cm}%
for~all~k$ \\ 
$\frac{d\beta }{dt}=\frac{\partial L}{\partial t}$%
\end{tabular}
\label{bilan}
\end{equation}

We recognize the equations of Lagrangian mechanics: $L$ being the
Lagrangian, the $(\alpha _{k})$ the generalized momentum, $\beta $ the
opposite of the Hamiltonian and $I$ the action. But this can not be correct
because, for instance, the dimension of action is $Action=Energy\times Time$
while $I$ is dimensionless. However, $K$ being an appropriate constant of
dimension $Action$, we recover all lagrangian mechanics with the following
identifications:

\begin{equation*}
\begin{tabular}{lll}
$K~I=S$ & \hspace{2cm} & action \\ 
$K~\alpha _{k}=p_{k}$ & \hspace{2cm} & generalized momentum \\ 
$K~\beta =-H$ & \hspace{2cm} & $-$ Hamiltonian \\ 
$K~L=\tciLaplace $ & \hspace{2cm} & Lagrangian%
\end{tabular}%
\end{equation*}

which give

\begin{equation}
\begin{tabular}{l}
$\tciLaplace ((q_{k}),(\overset{.}{q}_{k}),t)=\dsum\limits_{k}p_{k}\overset{.%
}{q}_{k}-H$ \\ 
$dS=\dsum\limits_{k}p_{k}dq_{k}-Hdt$ $=\tciLaplace dt$ \\ 
$\frac{dp_{k}}{dt}=\frac{\partial \tciLaplace }{\partial q_{k}}\hspace{3cm}%
for~all~k$ \\ 
$\frac{d(-H)}{dt}=\frac{\partial \tciLaplace }{\partial t}$%
\end{tabular}%
\end{equation}

The least information principle becomes the least action principle. The
action

$\hspace{3cm}S=\dint\limits_{t_{0}}^{t_{1}}\tciLaplace (t)dt$

is stationary. For any variations $\delta A_{k}(t)$ such that $\delta
A_{k}(t_{0})=\delta A_{k}(t_{1})=0$ for all $k$, we have $\delta S=0$.

\bigskip

Equations (\ref{dalpha}) and (\ref{dbeta}) become:

\begin{equation*}
\begin{tabular}{l}
$\frac{dp_{k}}{dt}=\frac{\partial \tciLaplace }{\partial q_{k}}\hspace{3cm}%
for~all~k$ \\ 
$\frac{d(-H)}{dt}=\frac{\partial \tciLaplace }{\partial t}$%
\end{tabular}%
\end{equation*}

which is Noether's theorem\cite{Noether}. The generalized momentum is
conserved if $\tciLaplace $ does not explicitly depend on the associated
generalized coordinate. Energy is conserved if $\tciLaplace $ does not
explicitly depend on time.

\section{Comments and open questions}

The relation between the Lagrangian $\tciLaplace $ and the Hamiltonian $H$
appears naturally in the Maxent formalism. However, while one is the
Legendre transform of the other, they do not play roles similar to $I$ and $%
\ln (Z)$. In fact, the partition function $Z$ does not play any particular
role in our description (the properties of $\ln (Z)$ mirror the properties
of $I$\cite{Dewar}), and it is the presence of time which induces the
relation between $\tciLaplace $ and $H$.

One can also note that, fundamentally, $\tciLaplace $ satisfies the
equations of Cauchy-Riemann because $dI$ is an exact differential.

\bigskip

In the case of conservative forces, LAP is equivalent to Newton's law. This
means that Laplace's equations could have emerge without any hint of Newton,
if the work of Jaynes (1922-1998) had been known. Maxent as a fundamental
physical principle certainly has epistemological implications.

\bigskip

For a given physical path, there is a constant $K$ with dimension $Action$
such that%
\begin{equation*}
\begin{tabular}{lll}
$K~I=S$ & \hspace{2cm} & action \\ 
$K~\alpha _{k}=p_{k}$ & \hspace{2cm} & generalized momentum \\ 
$K~\beta =-H$ & \hspace{2cm} & $-$ Hamiltonian \\ 
$K~L=\tciLaplace $ & \hspace{2cm} & Lagrangian%
\end{tabular}%
\end{equation*}

But nothing a priori prevents the value of $K$ to be different in different
experiments. One can not calculate the values of $I$ and of the $(\alpha
_{k})$ for a given physical path, since the $(q_{k,i})$ are a priori unknown.

However, the $(q_{k,i})$ could eventually be known using quantum mechanics
(QM). To simplify, let us state that there is one single coordinate $x$. In
QM, a system is defined by its wave function, which is a function of space
and time $\varphi (x,t)$. We can identify a path $i$ of probability $p_{i}$
with the set of all $(x,t)$ such that $\left\vert \varphi (x,t)\right\vert
^{2}=p_{i}$ (of course, to be rigorous, the discrete set of paths $(i)$
first has to be replaced by a continuous set). Since the paths $i$ are
known, the $(x_{i}(\lambda ))$ and $(t_{i}(\lambda ))$ are also known if we
can parameterize all these paths with a common parameter $\lambda $.
Deducing the value of $K$ rests an open question.

\bigskip

Feynmann's path integral formulation of QM\cite{Feynmann} offers
similarities with our description. However, the two theories also present
fundamental differences. In particular, Feynmann assigns equal probabilities
to all paths, and does not average the action $S$, but the quantity $e^{iS/%
\overline{h}}$. A possible link has to be investigated.

\bigskip

An analogy between mechanics and thermodynamics has already been found\cite%
{Gaies}. It uses the formalism of differential forms, but the main results
can be obtain using Maxent. This analogy comes fundamentally from the fact
that Lagrangian mechanics and equilibrium thermodynamics can both be
described by a set of linear constraints as has been shown in this paper.

\section{Conclusion}

The least action principle (LAP) is a consequence of Maxent, provided that
the constraints concern the average coordinates and the average time of a
mechanical system. The simplicity of the demonstration and the high degree
of generality of Maxent explain why LAP is so general in mechanics.

This demonstration of LAP sheds a new light on the relationship between
thermodynamics and mechanics. It offers an opportunity to unify these two
branches of physics, with Maxent as a common basis.

\pagebreak

\section*{Annex 1}

This annex demonstrates classical results about Maxent distributions\cite%
{Jaynes}.

Some numbers $(A_{k,i})$ are given once and for all, and will always be
considered fixed. Some other numbers $(A_{k})$ are given and we wish to find
the probability law $(p_{i})$ satisfying the constraints:

\begin{equation}
\left\{ 
\begin{array}{l}
\dsum\limits_{i}p_{i}A_{k,i}=A_{k}\qquad for~all~k \\ 
\dsum\limits_{i}p_{i}=1%
\end{array}%
\right.
\end{equation}

The Maxent principle consists of choosing the distribution $(p_{i})$ which
maximizes the information

\begin{equation*}
I=-\dsum\limits_{i}p_{i}~ln(p_{i})
\end{equation*}

Using the method of Lagrange multipliers, let

\begin{equation*}
\pounds =I-\dsum\limits_{k}\alpha _{k}\dsum\limits_{i}p_{i}A_{k,i}-\gamma
\dsum\limits_{i}p_{i}
\end{equation*}

where $(\alpha _{k})$ and $\gamma $ are new variables called the Lagrange
multipliers.

The distribution $(p_{i})$ should satisfy

\begin{equation*}
\frac{\partial \pounds }{\partial p_{i}}=0=-\ln
(p_{i})-1-\dsum\limits_{k}\alpha _{k}A_{k,i}-\gamma \qquad \qquad \text{for
all }i
\end{equation*}

Calling%
\begin{equation}
\begin{tabular}{l}
$Z_{i}=e^{-\dsum\limits_{k}\alpha _{k}A_{k,i}}$ \\ 
$Z=\dsum\limits_{i}Z_{i}$%
\end{tabular}
\label{Z}
\end{equation}

we obtain

\begin{equation*}
p_{i}=\frac{Z_{i}}{Z}
\end{equation*}

and%
\begin{equation*}
-\ln (p_{i})=\dsum\limits_{k}\alpha _{k}A_{k,i}+\ln (Z)
\end{equation*}

Therefore%
\begin{equation}
I=\dsum\limits_{i}p_{i}(-\ln (p_{i}))=\dsum\limits_{k}\alpha _{k}A_{k}+\ln
(Z)  \label{I}
\end{equation}

Differentiating (\ref{Z}), we obtain:%
\begin{equation*}
\frac{\partial Z_{i}}{\partial \alpha _{k}}=-A_{k,i}Z_{i}=-A_{k,i}p_{i}Z
\end{equation*}

\begin{equation*}
\frac{\partial Z}{\partial \alpha _{k}}=\dsum\limits_{i}\frac{\partial Z_{i}%
}{\partial \alpha _{k}}=-A_{k}Z
\end{equation*}

that we write:%
\begin{equation}
\frac{\partial \ln (Z)}{\partial \alpha _{k}}=-A_{k}\hspace{3cm}\text{for
all }k  \label{dlnZ}
\end{equation}

The $(\alpha _{k})$ can be found by solving the $k_{\max }$ equations of
this last system. $\ln (Z)$ is an exact differential and can be written:

\begin{equation*}
d(\ln (Z))=-\dsum\limits_{k}A_{k}d\alpha _{k}
\end{equation*}

Using equation (\ref{I}), we can now find an expression for $dI$. Since the
quantities $(A_{k})$ are independent, $\frac{\partial A_{k}}{\partial A_{l}}%
=\delta _{k,l}$ ($\delta _{k,l}$ is Kronecker symbol) and:%
\begin{equation*}
\frac{\partial I}{\partial A_{l}}=\dsum\limits_{k}\frac{\partial \alpha _{k}%
}{\partial A_{l}}A_{k}+\alpha _{l}+\frac{\partial \ln Z}{\partial A_{l}}
\end{equation*}

Since $Z$ is a function of the $\alpha _{k}$,

\begin{equation*}
\frac{\partial \ln Z}{\partial A_{l}}=\dsum\limits_{k}\frac{\partial \ln Z}{%
\partial \alpha _{k}}\frac{\partial \alpha _{k}}{\partial A_{l}}
\end{equation*}

and using (\ref{dlnZ}):

\begin{equation*}
\frac{\partial \ln Z}{\partial A_{l}}=\dsum\limits_{k}-A_{k}\frac{\partial
\alpha _{k}}{\partial A_{l}}
\end{equation*}

We finally obtain:

\begin{equation*}
\frac{\partial I}{\partial A_{k}}=\alpha _{k}\hspace{3cm}\text{for all }k
\end{equation*}

$I$ is a function of the $(A_{k})$:

\begin{equation*}
dI=\dsum\limits_{k}\alpha _{k}dA_{k}
\end{equation*}

$dI$ is an exact differential:%
\begin{equation*}
\frac{\partial ^{2}S}{\partial A_{k}\partial A_{l}}=\frac{\partial ^{2}S}{%
\partial A_{l}\partial A_{k}}
\end{equation*}

therefore:%
\begin{equation*}
\frac{\partial \alpha _{k}}{\partial A_{l}}=\frac{\partial \alpha _{l}}{%
\partial A_{k}}\ \ \ \ for~all~k,l
\end{equation*}

\pagebreak

\section*{Annex 2}

We have, by definition of $L$ (equation (\ref{L})):

\begin{equation*}
\beta =L-\dsum\limits_{k}\alpha _{k}\overset{.}{A_{k}}
\end{equation*}

So

\begin{equation*}
\frac{d\beta }{dt}=\frac{dL}{dt}-\dsum\limits_{k}\left( \frac{d\alpha _{k}}{%
dt}\overset{.}{A_{k}}+\alpha _{k}\overset{..}{A_{k}}\right)
\end{equation*}

when

\begin{equation*}
\overset{..}{A_{k}}=\frac{d\overset{.}{A_{k}}}{dt}
\end{equation*}

Since

\begin{equation*}
\frac{dL}{dt}=\dsum\limits_{k}\left( \frac{\partial L}{\partial A_{k}}%
\overset{.}{A_{k}}+\frac{\partial L}{\partial \overset{.}{A_{k}}}\overset{..}%
{A_{k}}\right) +\frac{\partial L}{\partial t}
\end{equation*}

taking into account (\ref{dalpha}) and (\ref{dL0}), we obtain (\ref{dbeta}):

\begin{equation*}
\frac{d\beta }{dt}=\frac{\partial L}{\partial t}
\end{equation*}

\pagebreak

\noindent
 
\end{document}